\documentclass[twocolumn]{aastex63}

\usepackage{amsmath}
\usepackage{makecell}
\usepackage{savesym}
\savesymbol{tablenum}
\usepackage[multi-part-units=single, separate-uncertainty=true]{siunitx}
\restoresymbol{SIX}{tablenum}
\usepackage{longtable}

\usepackage{graphicx}
\usepackage[caption=false]{subfig}

\DeclareSIUnit \parsec {pc}
\DeclareSIUnit \mas {mas}
\DeclareSIUnit \pixel {px}
\DeclareSIUnit \year {yr}

\received{June 14, 2021}
\revised{July 22, 2021}
\accepted{July 26, 2021}

\submitjournal{PASP}

\shorttitle{Brown Dwarf Parallax}
\shortauthors{Nguyen, Ammons, et al.}

\usepackage{makecell}

\begin{document}

\title{Trigonometric Parallaxes of Two T Dwarfs with Keck and ShaneAO Astrometry}

\correspondingauthor{Jayke Nguyen}
\email{jsn001@ucsd.edu}

\author{Jayke Nguyen}
\affiliation{Lawrence Livermore National Laboratory, 7000 East Ave,
Livermore, CA 94550, USA}
\affiliation{Center for Astrophysics and Space Sciences, Department of Physics, University of California, San Diego, 9500 Gilman Dr., La Jolla, CA 92093, USA}

\author{S. Mark Ammons}
\affiliation{Lawrence Livermore National Laboratory, 7000 East Ave,
Livermore, CA 94550, USA}

\author{Kaitlin Dennison}
\affiliation{Stanford University, Department of Aeronautics and Astronautics, Durand Building, 496 Lomita Mall, Stanford, CA 94305-4035, USA}

\author{E. Victor Garcia}
\affiliation{kWh Analytics, 609 Mission Street, Fl 2, San Francisco, CA 94105, USA}

\author[0000-0001-9611-0009]{Jessica R. Lu}
\affiliation{Astronomy Department, University of California, Berkeley, CA 94720, USA}

\author{Stephen McMillan}
\affiliation{Minnesota Institute for Astrophysics, University of Minnesota, 116 Church Street SE, Minneapolis, MN 55455, USA}

\author{Maissa Salama}
\affiliation{Institute for Astronomy, University of Hawai'i at Mānoa, Hilo, HI 96720-2700, USA}


\newcommand{\wiseBParallax}{$\pi = \SI{73.5\pm9.2}{\mas}$}
\newcommand{\wiseBParallaxNum}{$\SI{73.5\pm9.2}{\mas}$}
\newcommand{\wiseBProperMotionRA}{$\mu_\alpha = \SI{117.9\pm2.9}{\mas\per\year}$}
\newcommand{\wiseBProperMotionDec}{$\mu_\delta = \SI{403.4\pm2.2}{\mas\per\year}$}
\newcommand{\wiseBTableRow}{WISE1901 & $285.2759846(33)$ & $47.3056997(22)$ & $\phantom{-}117.9\pm2.9$ & $\phantom{-}403.4\pm2.2$ & $73.5\pm9.2\phantom{0}$ & $1.152$ \\}

\newcommand{\wiseCParallax}{$\pi = \SI{70.1\pm6.7}{\mas}$}
\newcommand{\wiseCParallaxNum}{$\SI{70.1\pm6.7}{\mas}$}
\newcommand{\wiseCProperMotionRA}{$\mu_\alpha = \SI{-166.7\pm2.9}{\mas\per\year}$}
\newcommand{\wiseCProperMotionDec}{$\mu_\delta = \SI{-463.1\pm3.5}{\mas\per\year}$}
\newcommand{\wiseCTableRow}{WISE2154 & $328.6367953(34)$ & $59.7032306(22)$ & $-166.7\pm2.9$ & $-463.1\pm3.5$ & $70.1\pm6.7\phantom{0}$ & $1.338$ \\}

\begin{abstract}

We present trigonometric parallax and proper motion measurements for two T-type brown dwarfs. We derive our measurements from infrared laser guide star adaptive optics observations spanning five years from the ShaneAO/SHARCS and NIRC2/medium-cam instruments on the Shane and Keck telescopes, respectively. To improve our astrometric precision, we measure and apply a distortion correction to our fields for both instruments. We also transform the Keck and ShaneAO astrometric reference frames onto the ICRS using five-parameter parallax and proper motion solutions for background reference stars from Gaia DR2. Fitting for parallax and proper motion, we measure parallaxes of \wiseBParallaxNum{} and \wiseCParallaxNum{} for WISEJ19010703+47181688 (WISE1901) and WISEJ21543294+59421370 (WISE2154), respectively. We utilize Monte Carlo methods to estimate the error in our sparse field methods, taking into account overfitting and differential atmospheric refraction. Comparing to previous measurements in the literature, all of our parallax and proper motion values fall within $2\sigma$ of the published measurements, and 4 of 6 measurements are within $1\sigma$. These data are among the first parallax measurements of these T dwarfs and serve as precise measurements for calibrating stellar formation models. These two objects are the first results of an ongoing survey of T dwarfs with Keck/NIRC2 and the Shane Adaptive Optics system at Lick Observatory.

\end{abstract}


\section{Introduction} \label{sec:intro}

A fundamental parameter of a stellar object is its distance from Earth. Historically, astronomers have constrained the distances of nearby stars with a variety of techniques, including the trigonometric parallax. Parallax measurements of stars can in turn be used to derive absolute magnitudes, allowing stars to be placed on color-magnitude diagrams and compared to theoretical star formation tracks. Brown dwarfs are substellar objects that were first theoretically devised as the lowest mass products of stellar formation processes \citep{Kumar_1962, Hayashi_1963}. They straddle the boundary between fusion-powered M dwarfs and giant planets. Brown dwarfs, particularly T and Y dwarfs, remain elusive in observational surveys due to their low luminosities at visible wavelengths. 

The advent of all-sky infrared surveys like the Sloan Digital Sky Survey (SDSS), Two Micron All Sky Survey (2MASS), Wide-field Infrared Survey Explorer (WISE), and UKIRT Infrared Deep Sky Survey (UKIDSS) has led to the discovery of hundreds of brown dwarfs in our stellar neighborhood \citep{Eisenstein_2011, Skrutskie_2006, Davy_Kirkpatrick_2011, Lawrence_2007}. However, only a small percentage of these objects have trigonometric parallax measurements. The dedicated astrometric survey Gaia \citep{Gaia_2016, Gaia_2018} has derived parallax and proper motions for millions of stars \citep{Lindegren_2018} but Gaia's astrometric precision suffers for faint T dwarfs that fall below its visible band sensitivity limit. Therefore, targeted observations at infrared wavelengths are required to derive parallaxes for the faintest of T dwarfs.

Current ground based laser guide star adaptive optics (LGSAO) systems provide relative astrometric precision of milliarcseconds or less \citep{Lu_2010, Ammons_2014, Neichel_2014} which can be used to derive trigonometric parallax of field brown dwarfs at similar levels of precision \citep{Ammons_2016, Garcia_2017}. The main drawback of LGSAO systems is that they are limited in their field of view to the anisokinetic angle of the site, typically \SI{20}{\arcsec} to \SI{40}{\arcsec}, in infrared wavelengths. Observing a single object only requires a few minutes of observation time per epoch, and many objects may be observed in a single night even with limited time. Nonetheless, this method of astrometry is a precise and reliable way to measure parallax and proper motion of faint T dwarfs.

We have selected a sample of 63 T dwarfs from the SDSS, 2MASS, WISE, and UKIDSS infrared surveys with nearby tip/tilt stars of brightness $R < 17$ and at least 3 background reference stars in a \SI{33}{\arcsec} diameter field of view. These brown dwarfs are currently being monitored astrometrically with the LGSAO systems at the 3-m Shane telescope at Lick Observatory \citep{Gavel_2014} and the 10-m Keck II telescope at Keck Observatory \citep{Wizinowich_2000, Wizinowich_2006, Max_1995}.

In this paper we present parallax and proper motion constraints for WISE1901 and WISE2154, the first two objects in our survey that have accumulated at least six epochs. We make use of a novel astrometric calibration method that works in sparse star fields. To validate our results, we compare our trigonometric parallax values to existing literature values \citep{Theissen_2018, Kirkpatrick_2020}.

Section \ref{sec:samp} describes the sample selection and survey strategy. Section \ref{sec:data} outlines the data reduction process for Shane and Keck data, including plate scale measurements for SHARCS and NIRC2. Section \ref{sec:astr} outlines stellar position finding using \texttt{Starfinder}, the alignment procedure, and the final parallax and proper motion fitting routine. Section \ref{sec:err} presents our error analysis using M92 data and Monte Carlo methods. Section \ref{sec:res} presents our measurements of the position, proper motion, and parallax for WISE1901 and WISE2154. Lastly, Section \ref{sec:conc} discusses our methodology and results in the context of astrometry and presents future prospects using our sparse-reference-star method.

\section{Sample and Survey Construction} \label{sec:samp}
\subsection{Laser Guide Star Adaptive Optics}
Using LGSAO at infrared wavelengths is advantageous for parallax measurements for one primary reason: the decreased point spread function (PSF) width compared to seeing-limited PSFs increases the signal-to-noise ratio (SNR) of detection against a bright infrared background. When faint stellar sources such as brown dwarfs are imaged without adaptive optics, astrometric errors due to SNR may be higher than the atmospheric differential tip/tilt jitter (DTTJ). With AO, these terms are reduced and, depending on the geometry of the reference stars constellation, astrometric errors can hit the systematic noise floor of the instrument within minutes of exposure time.

As mentioned previously, a limitation of LGSAO systems is the field of view as set by the anisokinetic angle, typically \SI{20}{\arcsec} - \SI{40}{\arcsec} at infrared wavelengths. This limited field of view occasionally fails to provide an adequate number of background reference stars. For this reason, the larger survey is limited to L and T dwarfs that happen to have at least three reference stars within this patch as well as a tip/tilt star for LGSAO guiding within the vicinity. Due to the dependence of astrometric DTTJ error on stellar separation, nearby reference stars are more stable than distant references, and only a few reference stars are needed to establish a local reference frame for position and rotation. In addition, all of our reference stars have positions and proper motions constrained in the ICRS system by the Gaia mission, which makes it possible to place our LGSAO observations from multiple observatories onto the same astrometric reference frame. We test the precision of this method and the implications for the errors on our parallax and proper motion with a Monte Carlo experiment in M92 in Section \ref{subsec:syserr}.

\subsection{Target Selection Criteria}
The larger sample was selected from the SDSS, 2MASS, WISE, and UKIDSS brown dwarf surveys. The criteria were: (1) spectral type later than L1; (2) at least three background reference stars with $R < 20$ in the USNO A2.0 catalog \citep{Monet_1998} within a diameter of \SI{33}{\arcsec} (within at least \SI{0.5}{\arcsec} of the edge, to ensure the PSF halo is entirely within the frame); and (3) a $R < 17$ tip/tilt star within \SI{60}{\arcsec} of the brown dwarf. The two targets presented here (see Table \ref{tab:bd_type_phot}), all selected from the WISE survey, are the two in our larger sample with the largest number of observations. 

Since the laser guide star was only used during bright time at both observatories, the observing cadence is roughly monthly with eight observing runs allocated per year. Lick Observatory is prone to poorer weather during the winter and spring months, so observations were typically only successful during the summer and early autumn time frames. This cadence presents a challenge for parallax measurements which ideally are sampled evenly throughout the year. To mitigate this effect, we make use of Keck LGS data as well which is less affected by seasonal weather. In addition, we obtained observations over a longer time baseline than is typical for parallax measurements (4 to 5 years, rather than the standard 1 or 2) to improve the SNR on the parallax and proper motion parameters. 

\section{Data} \label{sec:data}

\begin{deluxetable*}{ccccccccc}
\tabletypesize{\scriptsize}
\tablecaption{Brown Dwarf Type and Photometric Data}
\tablehead{
\colhead{Object} & \colhead{SpT} & \colhead{J-H} & \colhead{H-K$_s$} & \colhead{J-K$_s$} & \colhead{J} & \colhead{H} & \colhead{W1-W2} & \colhead{W2}
}
\startdata
WISE1901 & T5$^{(3)}$ & $0.39 \pm 0.12^{(3)}$ & $-0.17 \pm 0.07^{(3)}$ & \ldots & $15.86 \pm 0.07^{(3)}$ & \ldots & \ldots & \ldots \\
WISE2154 & T5$^{(1)}$ & \ldots & \ldots & \ldots & $15.66^{(1)*}$ & $15.77 \pm 0.17^{(2)}$ & $2.14 \pm 0.08^{(2)}$ & $13.51 \pm 0.03^{(2)}$ \\
\enddata
\tablecomments{
References. (1) \cite{Looper_2007}, (2) \cite{Davy_Kirkpatrick_2012}, (3) \cite{Burgasser_2004} \\ $^*$ 2MASS J band
}
\label{tab:bd_type_phot}
\end{deluxetable*}

\begin{deluxetable*}{ccccccc}
\tabletypesize{\normalsize}
\tablecaption{Gaia Reference Stars}
\tablehead{
\colhead{Label} & \colhead{J2000 RA} & \colhead{J2000 Dec} & \colhead{PM RA} & \colhead{PM Dec} & \colhead{Parallax} & \colhead{Gaia Source ID} \\
\colhead{} & \colhead{hr} & \colhead{deg} & \colhead{mas/yr} & \colhead{mas/yr} & \colhead{mas} & \colhead{}
}
\decimals
\startdata
\multicolumn{7}{c}{\textbf{WISE1901}} \\
\hline
\multicolumn{7}{c}{\textbf{Shane Reference Stars}} \\
1901S0 & 19 01 07.012 & +47 18 16.772 & \phantom{-}0.196 $\pm$ 0.190 & -3.984 $\pm$ 0.184 & 0.318 $\pm$ 0.093 & 2131382365365031040 \\
1901S2 & 19 01 06.219 & +47 18 01.525 & -1.335 $\pm$ 0.239 & -3.908 $\pm$ 0.234 & 0.736 $\pm$ 0.116 & 2131383125572944768 \\
1901S3 & 19 01 06.022 & +47 17 59.638 & -1.113 $\pm$ 0.038 & \phantom{-}6.649 $\pm$ 0.039 & 1.566 $\pm$ 0.019 & 2131383125576327552 \\
\\
\multicolumn{7}{c}{\textbf{Keck Reference Stars}} \\
1901K1 & 19 01 07.012 & +47 18 16.772 & \phantom{-}0.196 $\pm$ 0.190 & -3.984 $\pm$ 0.184 & 0.318 $\pm$ 0.093 & 2131382365365031040 \\
1901K2 & 19 01 06.817 & +47 18 03.618 & \phantom{-}0.685 $\pm$ 0.662 & -4.982 $\pm$ 0.632 & 0.394 $\pm$ 0.322 & 2131382365365026048 \\
1901K3 & 19 01 06.219 & +47 18 01.525 & -1.335 $\pm$ 0.239 & -3.908 $\pm$ 0.234 & 0.736 $\pm$ 0.116 & 2131383125572944768 \\
\hline
\multicolumn{7}{c}{\textbf{WISE2154}} \\
\hline
\multicolumn{7}{c}{\textbf{Shane Reference Stars}} \\
2154S1 & 21 54 32.709 & +59 42 03.752 & -2.603 $\pm$ 0.439 & -2.699 $\pm$ 0.425 & \phantom{-}0.670 $\pm$ 0.205 & 2202738436624981376 \\
2154S2 & 21 54 31.314 & +59 42 01.835 & -1.803 $\pm$ 1.294 & -1.857 $\pm$ 0.961 & \phantom{-}0.773 $\pm$ 0.497 & 2202738436624980608 \\
2154S3 & 21 54 30.856 & +59 42 10.874 & -2.132 $\pm$ 0.773 & -2.841 $\pm$ 0.631 & -0.425 $\pm$ 0.343 & 2202739192539229056 \\
\\
\multicolumn{7}{c}{\textbf{Keck Reference Stars}} \\
2154K1 & 21 54 32.709 & +59 42 03.752 & -2.603 $\pm$ 0.439 & -2.699 $\pm$ 0.425 & \phantom{-}0.670 $\pm$ 0.205 & 2202738436624981376 \\
2154K2 & 21 54 31.314 & +59 42 01.835 & -1.803 $\pm$ 1.294 & -1.857 $\pm$ 0.961 & \phantom{-}0.773 $\pm$ 0.497 & 2202738436624980608 \\
2154K3 & 21 54 30.856 & +59 42 10.874 & -2.132 $\pm$ 0.773 & -2.841 $\pm$ 0.631 & -0.425 $\pm$ 0.343 & 2202739192539229056 \\
\enddata
\tablecomments{Only WISE1901 uses different reference stars than our Shane data. For consistency between targets, matching reference stars between Shane and Keck data are duplicated in the table. Negative parallaxes are technically unphysical, but allowed in Gaia data \citep{Luri_2018}. In our methodology, we treat negative parallaxes as zero.}
\label{tab:gaia_ref}
\end{deluxetable*}

\subsection{Shane Data} \label{subsec:shane}

\begin{figure}[ht] 
\centering
\includegraphics[width=0.8\columnwidth]{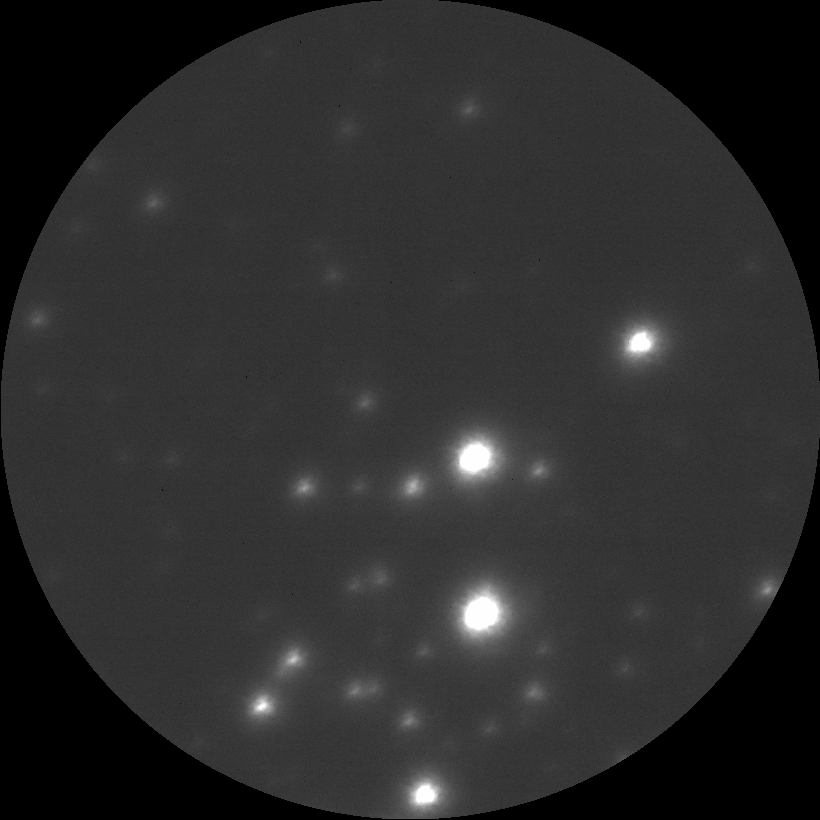}
\caption{Example \SI{1020}{\pixel} by \SI{1020}{\pixel} cropped frame from 2014 May 6 of processed M92 data from ShaneAO. This frame has been processed to remove bad pixels and a circular crop on the science field is applied. The circle has a radius of \SI{510}{\pixel} and the sensor has a plate scale of \SI{32.812}{\mas\per\pixel}, \SI{32.806}{\mas\per\pixel} in $x$ and $y$ respectively.}
\label{fig:m92shane}
\end{figure}

We observed brown dwarf targets using the SHARCS camera with ShaneAO on the Shane telescope from 2014 November 8 to 2019 August 17. The SHARCS camera uses a circular subsection of a cooled, engineering grade Teledyne HAWAII-2RG chip about 1020 pixels in diameter, with a \SI{33}{\arcsec} field of view. All ShaneAO brown dwarf science frames were taken in H band and have a total exposure time of \SI{120}{\second}. All observing sequences are undithered and we attempt to return the field to the same pixel in-between epochs by using the same offsets from the tip/tilt star. Within epochs we see $\sim$\SI{1}{\pixel} of jitter due to errors in telescope pointing over time. Between epochs, we see $\sim$\SI{10}{\pixel} of unintentional shifts due to stellar proper motions and variations in mechanical tip/tilt mirror response, such as actuator hysteresis. An example of a SHARCS M92 frame is shown in Figure \ref{fig:m92shane}. The flat-field images were constructed by median combining twilight flats when available and dome flats otherwise. Dark frames were also subtracted using frames matching the exposure time of our science targets. The Gaia reference stars used for aligning Shane data are found on Table \ref{tab:gaia_ref}. A list of targets and the number of frames used are listed on Tables \ref{tab:m92obs} and \ref{tab:obs}.

\begin{figure}[ht]
\centering
\includegraphics[width=0.8\columnwidth]{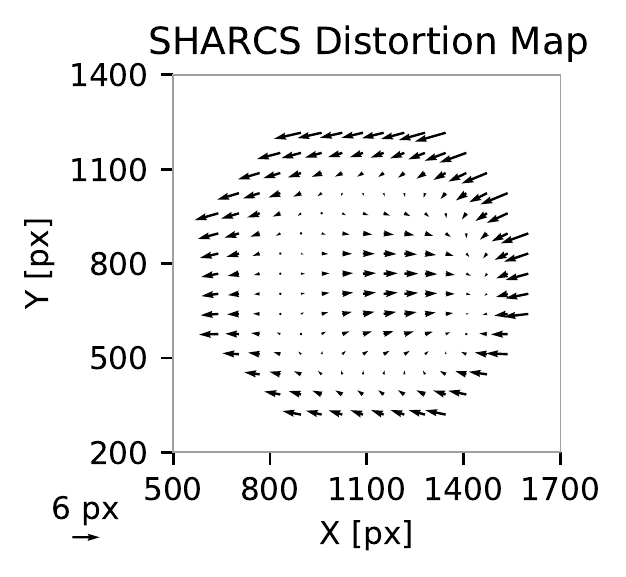}
\caption{Vector field of the SHARCS distortion map across the science field subset of the image sensor in absolute pixel units. The pixel coordinates range from 0 to 2048 in $x$ and $y$ but only a circular subsection is used. The circle is centered at roughly $(\SI{1108}{\pixel},\SI{769}{\pixel})$ with a radius of $\SI{510}{\pixel}$. Note that the distortion map is large in magnitude and poorly constrained at the edge of the bounding circle since there are few stars used in the fit near the edge.}
\label{fig:sharcsdist}
\end{figure}

For all stellar positions estimated from SHARCS data, we apply a fourth order polynomial distortion map derived from M92 data taken on 2014 August 12. The resulting distortion map is shown in Figure \ref{fig:sharcsdist}. We directly map matching stars in SHARCS M92 observations to Hubble ACS M92 observations \citep{Yelda_2010}. A gnomonic projection is not necessary since we do not transform into a spherical coordinate system at this step. The distortion map utilizes 205 stars and fits only the nonlinear components of the combined optical and physical sensor distortion. The distortion map is well defined over the circular subsection, but is less accurate near the edges due to a declining density of well-detected stars.

All of our Shane data, except our M92 data, were taken with the laser guide star. Although ShaneAO was initially near-diffraction-limited in 2014, starting in 2016 the LGS return flux declined, resulting in a much broader PSF and decreased precision. To combat this, we implement a full width half maximum (FWHM) cut, removing data above the 75th percentile. As mentioned previously, seeing conditions varied significantly and were generally fair, but some data are affected by poor seeing conditions.

\subsection{Keck Data} \label{subsec:keck}

\begin{figure}[ht]
\centering
\includegraphics[width=0.8\columnwidth]{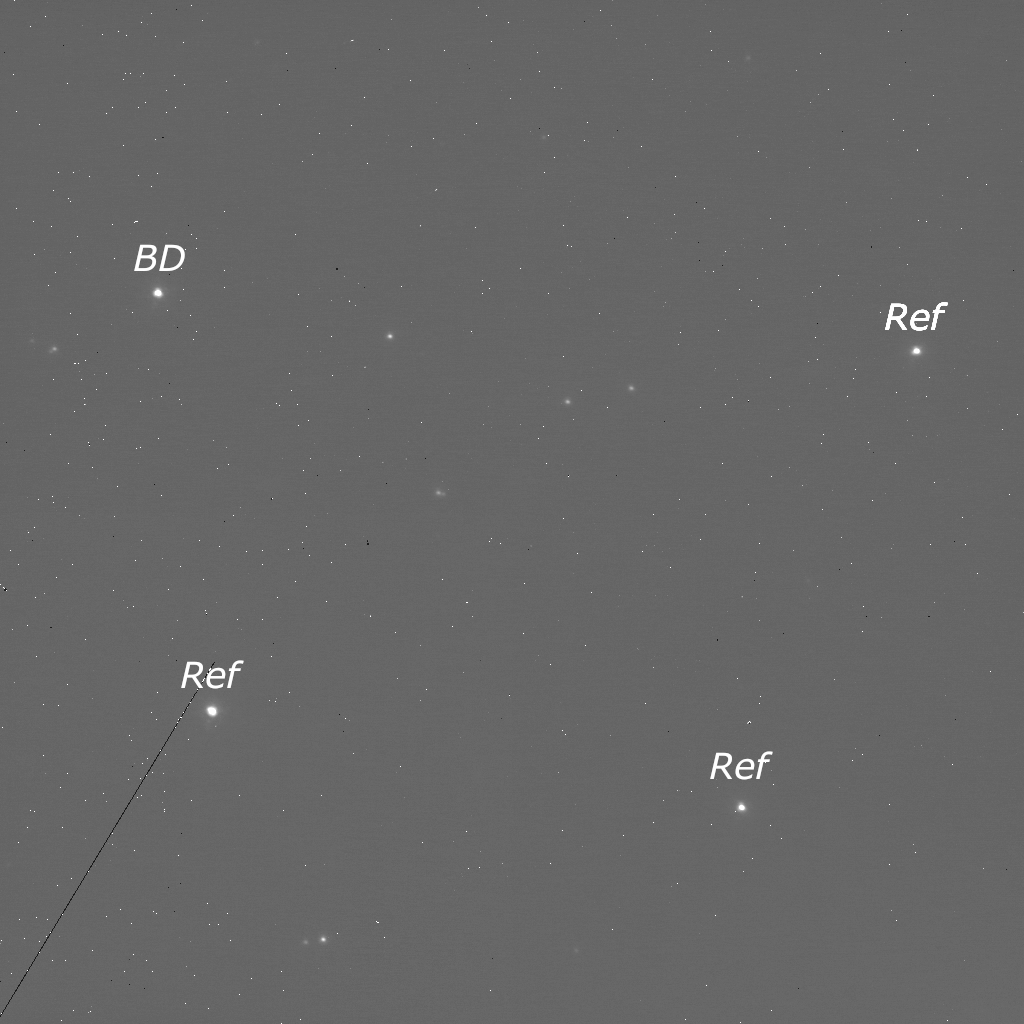}
\caption{Example \SI{1024}{\pixel} by \SI{1024}{\pixel} frame of WISE2154 from Keck taken on 2014 September 8. The fainter unlabeled stars with low SNR are not visible in Shane images and are thus unsuitable reference stars. We measure the plate scale of NIRC2 medium cam in $x$ and $y$ to be \SI{19.885}{\mas\per\pixel} and \SI{19.788}{\mas\per\pixel}, respectively, pre-bench realignment and a plate scale of \SI{19.849}{\mas\per\pixel} post-bench realignment, where the bench realignment occurred on 2015 April 13. The reference stars and brown dwarf are labeled ``Ref'' and ``BD'' in the image. Note that NIRC2 has very few bad pixels as compared to SHARCS. The field was also positioned to ensure our data are not impacted by the crack in the bottom left of the sensor.}
\label{fig:exframekeck}
\end{figure}

We obtained images of our brown dwarf target fields using the NIRC2 medium camera with the LGSAO system at the 10-m Keck II Observatory. We use data from 2014 August 6, 2014 September 8, 2015 June 4, and 2015 June 30. NIRC2 medium camera has a field of view of \SI{20}{\arcsec} with an Aladin-3 $1024\times1024$ pixel array. All data was taken with a \SI{60}{\second} exposure time in the Ks band. Like the Shane data, Keck frames were not dithered. Within epochs we see $\sim$\SI{0.5}{\pixel} of jitter due to errors in telescope pointing over time. Between epochs, we see $\sim$\SI{3}{\pixel} of unintentional shifts due to stellar proper motions and variations in mechanical tip/tilt mirror response, such as actuator hysteresis. Images are flat fielded using dome flats or twilight flats and dark subtracted using standard techniques. An example of a NIRC2 brown dwarf science frame is shown in Figure \ref{fig:exframekeck}. The NIRC2 sensor has very few bad pixels and did not require any bad pixel treatment. Compared to SHARCS data, the diffraction-limited NIRC2 data are far more precise and seeing conditions were excellent on all four epochs. The Gaia reference stars used for aligning Keck data are found on Table \ref{tab:gaia_ref}. A list of targets and number of frames used are listed on Tables \ref{tab:m92obs} and \ref{tab:obs}.

\begin{figure}[ht]
\centering
\includegraphics[width=0.8\columnwidth]{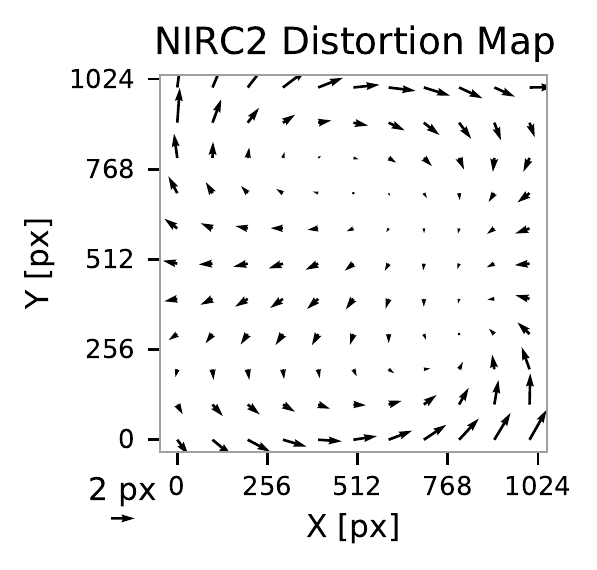}
\caption{NIRC2 medium camera distortion map across the $\SI{20}{\arcsec} \times \SI{20}{\arcsec}$ field of view, fit using a sixth-order polynomial. Since the medium camera on NIRC2 is not often used, the distortion map was not previously measured and had to be directly measured from M92 observations from 2015 April 4.}
\label{fig:nirc2dist}
\end{figure}

\begin{figure}[ht]
\centering
\includegraphics[width=0.8\columnwidth]{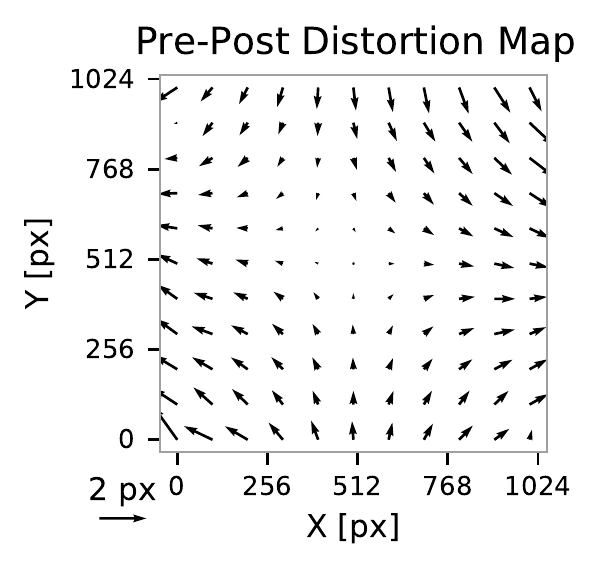}
\caption{NIRC2 distortion map difference between pre- and post- bench realignment across the whole sensor. The arrow scale is increased by a factor of two relative to Figure \ref{fig:nirc2dist}. The bench realignment corrected this, resulting in the difference distortion map showing an elongation of the x component and compression the y component of the distortion correction.}
\label{fig:nirc2distdiff}
\end{figure}

To create a distortion map for the Keck data, we fit a sixth-order polynomial to 919 stars in M92 data from 2015 April 4 using the medium camera configuration of NIRC2 (Figure \ref{fig:nirc2dist}). However, on 2015 April 13, a bench realignment of the Keck II telescope optics was performed, correcting a y-axis PSF elongation, changing the optical distortion pattern on the sensor. Our first two epochs occur before this date, and our last two epochs occur after. We were not able to collect M92 data post-bench realignment but the a distortion map was measured before and after this date using the narrow camera on NIRC2 \citep{Yelda_2010, Service_2016}. The pre- distortion map is a bivariate b-spline fit, while the post- distortion map is a sixth-order polynomial fit. Changing cameras is effectively a change in the optical system and therefore the narrow camera distortion map cannot be directly applied to data taken on the medium camera. However, using the difference between the pre- and post- distortion map we measure the change in distortion after the bench realignment and apply this change to our pre-bench realignment distortion map, effectively creating a distortion map for post-bench realignment data on medium camera. The difference distortion map is shown in Figure \ref{fig:nirc2distdiff} and is visually a good match to \cite{Service_2016}.

\subsection{SHARCS Plate Scale Measurements} \label{subsec:sharcsplate}

\begin{figure}[ht] 
\centering
\includegraphics[width=0.8\columnwidth]{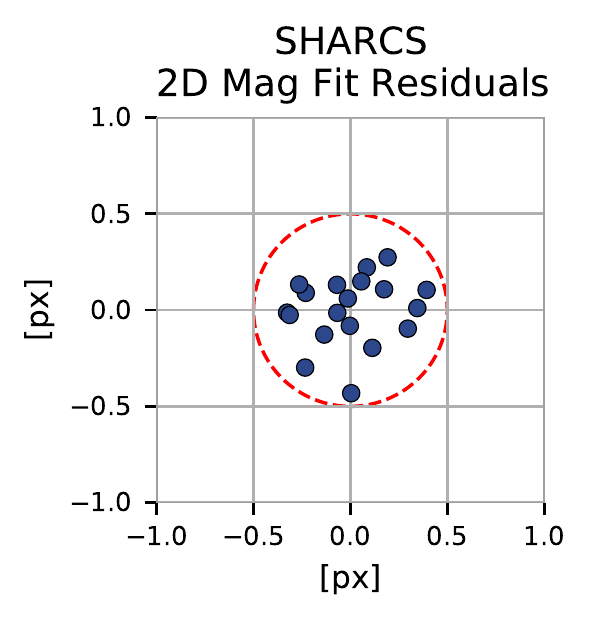}
\caption{Residuals of the SHARCS plate scale linear transform fit. The standard deviations of the residuals in $x$ and $y$ are $(\sigma_x,\sigma_y) = (\SI{0.213}{\pixel}, \SI{0.172}{\pixel})$. The distortion map was applied to stellar positions measured from M92 data collected from the Shane telescope on 2014 August 12 and compared to Hubble ACS astrometric measurements \citep{Yelda_2010}}
\label{fig:sharcsps}
\end{figure}

Measurements of the plate scale were made by cross referencing distortion-corrected M92 data against distortion-corrected Hubble ACS data \citep{Yelda_2010}. Since we compare two projected systems, a gnomonic projection is not needed and we may fit a linear transformation between the two systems directly. Using observations of M92 collected on 2014 August 12, we use 19 reference stars to measure the plate scale by fitting a five-parameter linear transformation (x/y translation, rotation, and x/y magnification) to the values. The resulting plate scale is \SI{32.812}{\mas\per\pixel}, \SI{32.806}{\mas\per\pixel} in $x$ and $y$ respectively, which we use as a fixed value in all transformations. Figure \ref{fig:sharcsps} shows the residuals from that linear transformation. The standard deviations of the residuals in $x$ and $y$ are $(\sigma_x,\sigma_y) = (\SI{0.213}{\pixel}, \SI{0.172}{\pixel})$.

\subsection{NIRC2 Plate Scale Measurements} \label{subsec:nirc2plate}

\begin{figure}[ht]
\centering
\includegraphics[width=0.8\columnwidth]{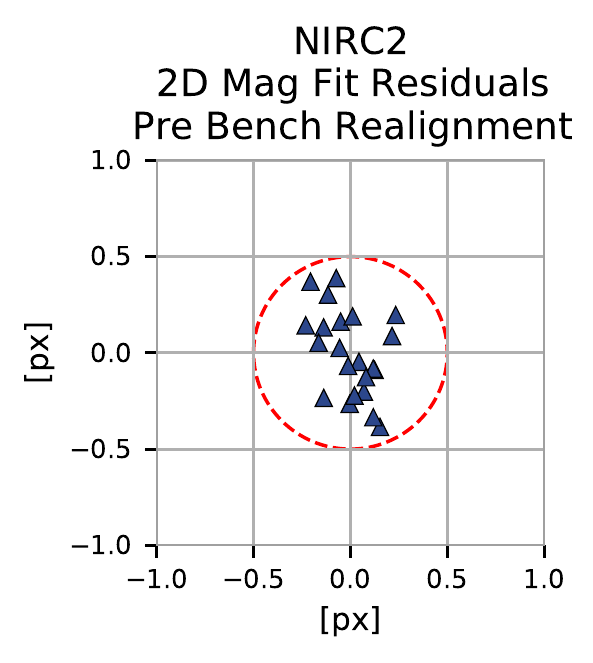}
\caption{Residuals of the pre-bench realignment NIRC2 plate scale fit using M92 data from 2014 August 5. The standard deviations of the residuals in $x$ and $y$ are $(\sigma_x,\sigma_y) = (\SI{0.129}{\pixel},\SI{0.217}{\pixel})$. We fit magnification in $x$ and $y$ since we expect an optical defect that preferentially elongates the PSF in the $x$ direction \citep{Service_2016}.}
\label{fig:nirc2preb}
\end{figure}

\begin{figure}[ht]
\centering
\includegraphics[width=0.8\columnwidth]{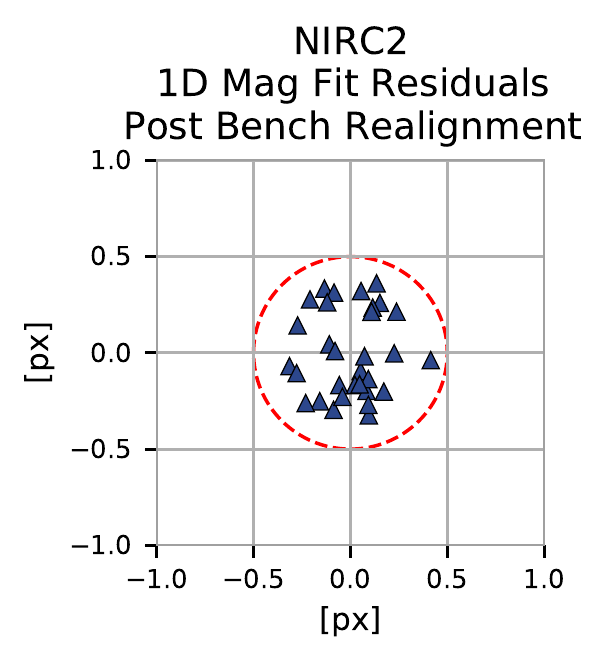}
\caption{Residuals of the post-bench realignment NIRC2 plate scale fit. The standard deviations of the residuals in $x$ and $y$ are $(\sigma_x,\sigma_y) = (\SI{0.167}{\pixel},\SI{0.219}{\pixel})$. We fit a single magnification to the plate scale since after the bench realignment the PSFs we obtain should now be very close to square \citep{Service_2016}. Fits to the two dimensional plate scale post-bench realignment data show a less than a $0.01\%$ difference between $x$ and $y$, confirming that a single plate scale measurement for both dimensions is sufficient.}
\label{fig:nirc2postb}
\end{figure}

The process for measuring the NIRC2 plate scale is almost exactly the same procedure as outlined in Section \ref{subsec:sharcsplate}. However, we know that prior to the bench realignment there is an expected y-axis elongation in NIRC2. Pre-bench realignment data is fit with a x/y translation, rotation, and x/y magnification, while post-bench realignment data is fit with a x/y translation, rotation, and single magnification. Additionally, we apply the difference distortion map to the post-bench realignment data. For the pre-bench realignment we use 22 data points and measure an $x$ and $y$ plate scale of \SI{19.885}{\mas\per\pixel} and \SI{19.788}{\mas\per\pixel}, respectively. For post-bench realignment data we use 31 data points and measure a single plate scale of \SI{19.849}{\mas\per\pixel}. Figures \ref{fig:nirc2preb} and \ref{fig:nirc2postb} shows the residuals of the pre- and post- bench realignment, respectively. The standard deviations of the pre- and post- bench realignment residuals in $x$ and $y$ are $(\sigma_x,\sigma_y) = (\SI{0.129}{\pixel},\SI{0.217}{\pixel})$ and $(\sigma_x,\sigma_y) = (\SI{0.167}{\pixel},\SI{0.219}{\pixel})$, respectively.

\begin{deluxetable*}{cccccc}
\label{tab:m92obs}
\tabletypesize{\normalsize}
\tablecaption{List of M92 Shane and Keck Observations}
\tablehead{
 \colhead{Date} & \colhead{Start Time} & \colhead{Instrument} & \colhead{Band} & \colhead{No. Frames} & \colhead{Integration Time} \\
 \colhead{UTC} & \colhead{UTC} & \colhead{} & \colhead{} & \colhead{} & \colhead{sec}
}
\startdata
2014-04-15 & 12:03:37.598 & SHARCS, Shane & H & 2 & 15\\
2014-05-07 & 09:37:33.111 & SHARCS, Shane & H & 9 & 15\\
2014-08-12 & 04:43:33.644 & SHARCS, Shane & H & 1 & 15\\
2014-08-05 & 08:20:36.24 & NIRC2, Keck & Ks & 1 & 10\\
2015-04-03 & 12:35:21.91 & NIRC2, Keck & Ks & 1 & 30\\
\enddata
\end{deluxetable*}

\section{Astrometry} \label{sec:astr}
\subsection{Overview}
We develop an coordinate transformation procedure to unite astrometry from both Shane and Keck onto the same astrometric system, the International Celestial Reference System (ICRS). All steps utilize positions measured from three reference stars in the field. As a first step, we align frames within epochs to the epoch mean with two parameter translation fits. Next, we place the multiple epochs onto the same reference frame using three parameter translation and rotation fits. We then derive a World Coordinate System (WCS) transform onto the ICRS using Gaia positions for reference stars. Lastly, we fit the five-parameter parallax and proper motion solutions to the brown dwarf positions on the ICRS. Astrometry derived from Shane and Keck are aligned separately until the final parallax and proper motion fit at which point they are both on the ICRS.
\begin{figure}[ht] 
\centering
\includegraphics[width=0.4\columnwidth]{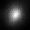}
\caption{An output template PSF of Shane WISE2154 data from \texttt{StarFinder}. The template PSF is stacked with other objects in the frame, smoothed, and normalized. Stacking all of the objects in the frame help minimize the bias introduced by bad pixel replacement.}
\label{fig:temppsf}
\end{figure}

\subsection{Stellar Position Finding} \label{subsec:stellarpos}
We use \texttt{StarFinder} \citep{Diolati_2000} to find the stellar position of each object in our science field. In each frame, a template PSF \SI{30}{\pixel} wide is estimated using all of the objects in the frame. Figure \ref{fig:temppsf} shows an example of a template PSF used to find stellar positions of objects in the frame. Each object used to create the template PSF is interpolated, allowing us to center with sub-pixel accuracy. However, the interpolation methods require a clean image free of bad pixels to determine the stellar position properly. Therefore, we use pixel value replacement methods to fix the bad pixels in the image. We use \texttt{astroscrappy}, a Python implementation of the L.A. Cosmic algorithm \citep{McCully_2018,van_Dokkum_2001} in addition to sigma clipping to replace bad pixels with the median value of their neighbors. Since all of our targets are relatively isolated, we also cropped the relevant objects in each frame and created a mosaiced image that would run through \texttt{StarFinder} faster. The initial crop parameters are used to recover the actual stellar positions on the frame. In the final stack of translated PSFs, we mask bad pixels, take the median pixel values across the entire image stack, and smooth with a variable box size median filter to create the master PSF for the frame. Each master PSF has at least three stacked sources per frame. This master PSF is then correlated with all the objects in the frame and the stellar position is determined by aligning the master PSF to each object.

Immediately after finding the stellar position, we apply the SHARCS or NIRC2 distortion map to the data. Each stellar position found in each frame is distortion-corrected.

\subsection{Within-Epoch Alignment} \label{subsec:within}

Within a given epoch, we have several frames from which we derive a single set of stellar position coordinates for all the unique objects in the frame. First, we take the mean stellar position of each object in the field, averaging the distortion-corrected stellar position of each object across frames, and use the averaged stellar position object to create a ``mean frame''. A simple two parameter translation is fit (Equation \ref{eq:trans}) and applied to each frame to shift the stellar positions as close as possible to the mean frame. A translation is sufficient as we expect telescope tracking errors to be small. To handle outliers, we iteratively shrink a circle centered on the average stellar position of each object in the frame, starting at \SI{100}{\mas} in radius and shrinking to \SI{20}{\mas} over \SI{1}{\mas} intervals, rejecting any frames with stellar positions outside of that radius. Any time a frame is rejected, the mean frame is remeasured and refitted. The set of stellar positions measured taking the average stellar positions of the translated frames is our final measurement for the epoch. We repeat this procedure for each epoch, producing a single final mean frame per epoch.

\begin{equation}
\begin{bmatrix}
x \\
y
\end{bmatrix}
+
\begin{bmatrix}
x_0 \\
y_0
\end{bmatrix}
=
\begin{bmatrix}
x_f \\
y_f
\end{bmatrix}
\label{eq:trans}
\end{equation}

\subsection{Epoch-to-Epoch Alignment} \label{subsec:e2e}
Using the resulting final mean frame from each epoch, we align the new set of epoch frames using a similar process as Section \ref{subsec:within}. Given the large temporal difference between epochs, we expect that there are much larger systematic errors that result from telescope pointing error, changes in the adaptive optics system, and differing atmospheric parameters in each observation. To account for small rotations in the fields between epochs, we fit an additional rotational parameter (Equation \ref{eq:transrot}). Additionally, since we expect the brown dwarf to have high proper motion, the brown dwarf is excluded from the alignment fit but included when the fit is applied. The brown dwarf is easily identified as the object that moves significantly between epochs relative to the other objects. 

\begin{equation}
\begin{bmatrix}
\cos{\theta} & -\sin{\theta} \\
\sin{\theta} & \cos{\theta}
\end{bmatrix}
\begin{bmatrix}
x \\
y
\end{bmatrix}
+
\begin{bmatrix}
x_0 \\
y_0
\end{bmatrix}
=
\begin{bmatrix}
x_f \\
y_f
\end{bmatrix}
\label{eq:transrot}
\end{equation}

\subsection{Epoch-to-Gaia Alignment} \label{subsec:e2g}
Lastly, we fit a gnomonic projection WCS transform to convert our epoch aligned data to the ICRS. The fit parameters in the WCS transform are \texttt{CRVAL1} (translation), \texttt{CRVAL2} (translation), and $\theta$ (rotation). \texttt{CRPIX1} and \texttt{CRPIX2} are assumed to be in the middle of the image, while the $x$ and $y$ plate scale, $k_x$ and $k_y$ (Equation \ref{eq:transrotmag}), are measured independently from M92 data. At minimum, we require that we have three reference stars for an accurate fit to Gaia. Knowing the relative shape of the background stars and the plate scale of both SHARCS and NIRC2, we are able to find the ICRS coordinates of the background stars in Gaia DR2. After removing the parallax and proper motion of the reference stars, we fit our data to the Gaia coordinates with a WCS transform. The resulting output yields the coordinates of the brown dwarf at each epoch in right ascension and declination in the ICRS.

\begin{equation} \label{eq:transrotmag}
\begin{bmatrix}
\cos{\theta} & -\sin{\theta} \\
\sin{\theta} & \cos{\theta}
\end{bmatrix}
\begin{bmatrix}
k_x & 0 \\
0 & k_y \\
\end{bmatrix}
\begin{bmatrix}
x \\
y
\end{bmatrix}
+
\begin{bmatrix}
x_0 \\
y_0
\end{bmatrix}
=
\begin{bmatrix}
x_f \\
y_f
\end{bmatrix}
\end{equation}

\subsection{Parallax Fit} \label{subsec:para}

The parallax and proper motion equation is given by \cite{Davy_Kirkpatrick_2011} and \cite{Urban_2013}:
\begin{equation}
\cos{\delta_1}(\alpha_i - \alpha_1) = \Delta_\alpha + \mu_\alpha (t_i
- t_1) + \pi \vec{R_i} \cdot \hat{W}
\label{eq:para_ra}
\end{equation}

\begin{equation}
(\delta_i - \delta_1) = \Delta_\delta + \mu_\delta (t_i - t_1) -
\pi \vec{R_i} \cdot \hat{N}
\label{eq:para_dec}
\end{equation}

Where the north-pointing and and west-pointing vectors are defined as:

$$
\hat{N} = -\hat{x} \cos{\alpha}\sin{\delta} - \hat{y}
\sin{\alpha}\sin{\delta} + \hat{z} \cos{\delta}
$$
$$
\hat{W} = \hat{x} \sin{\alpha} - \hat{y} \cos{\alpha}
$$

$\alpha$ is the right ascension, $\delta$ is the declination, $\Delta_\alpha$ is the right ascension offset, $\Delta_\delta$ is the declination offset, $\mu_\alpha$ is the right ascension proper motion, $\mu_\delta$ is the declination proper motion, $\vec{R}$ is the Earth's position vector, $t$ is the time, and $\pi$ is the parallax. Subscript $1$ denotes the initial epoch at time $t_1$ and subscript $i$ denotes a particular epoch at time $t_i$.

From the brown dwarf positions that are mapped to the ICRS, we fit Equations \ref{eq:para_ra} and \ref{eq:para_dec} to the data. As a result, we get a measurement on the parallax and proper motion of the brown dwarf target.

It is also important to note that Shane and Keck data are aligned separately until the final parallax and proper motion fit is performed. Only then are both the Keck and Shane data in the same ICRS coordinate system. 

\section{Error and Monte Carlo Analysis} \label{sec:err}

\subsection{Positional Error} \label{subsec:poserr}
As a first order estimate of our positional error, we compute the standard deviation in $x$ and $y$ of the reference star residuals after the epoch-to-epoch alignment step. In this computation we include residuals from all epochs, combining both Shane and Keck data. The resulting errors are: (\SI{6.733}{\mas}, \SI{7.510}{\mas}) and (\SI{5.338}{\mas}, \SI{2.963}{\mas}) for WISE1901 and WISE2154 respectively. This calculation is a simple estimate of the random portion of the positional error and will capture effects from varying PSF FWHM widths, low SNR, and errors due to bad pixel correction. To determine the true systematic error, a more rigorous analysis of our sparse-reference star method, capturing effects from overfitting as well as contributions from differential atmospheric refraction, needs to be performed.

\subsection{Systematic Error Due to Overfitting} \label{subsec:syserr}

With only a few reference stars per target, we risk overfitting our data if we use transformations with too many parameters. With three reference stars we have three parameters to fit and six data points measured per frame (each reference star has two data points, $x$ and $y$). Any additional fitting terms (e.g. magnification, skew) in the transformations could result in additional error due to overfitting. To model this, in our positional errors, we add an extra term in quadrature that captures the systematic overfitting error of our alignment process due to only using three reference stars. 

We estimate this systematic overfitting component by simulating sparse-field parallax and proper motion measurements with M92 cluster data taken over multiple epochs. Unlike our targets, M92 has many bright stars in the field, all with effectively zero parallax and high SNR. For any random selection of three ``reference'' stars in M92 and one mock ``brown dwarf'' we would expect to derive close to zero relative motion over two successive epochs using our sparse-reference star method for alignment. Any residuals we do detect would be mostly due to overfitting.

We start with two M92 frames acquired with ShaneAO on 2014 April 14 and nine frames on 2014 May 6, from which stellar positions have been measured with \texttt{StarFinder}. For each of 1820 Monte Carlo iterations, we randomly pick combinations of four stars. These combinations are analogous to having three reference stars and one virtual brown dwarf. We perform our alignment and fitting routine using the reference stars for alignment, stopping at the epoch-to-epoch step. By excluding the epoch-to-Gaia alignment, we ensure that this simulation accounts for sources of systematic error such as jitter between frames and unintentional dithering between epochs. We then measure the change in position of the mock brown dwarf between the two epochs. We then compute the standard deviation of these positional changes for all 1820 Monte Carlo iterations. We divide this standard deviation of the residuals by a factor of $\sqrt{2}$ as we are using a difference between two epochs to estimate the error for a single epoch. The resulting single-axis systematic errors from overfitting are ($\sigma_x$,$\sigma_y$) = (\SI{8.989}{\mas}, \SI{6.852}{\mas}) when using three reference stars.

\subsection{Differential Atmospheric Refraction} \label{subsec:dar}

Light from stars passing through the atmosphere is refracted by different amounts depending on the zenith angle at which it is observed and the spectrum of the star. This results in a difference in separation between two stars in the field due to the atmosphere, the differential atmospheric refraction (DAR). This phenomena is generally not a large effect, but becomes important at milliarcsecond level precision.

For each target, we compute the achromatic and chromatic differential atmospheric refraction (ADAR, CDAR) corrections for all frames according to the model defined by \cite{Gubler_1998}. The ADAR and CDAR are calculated using ambient temperature, pressure, humidity, and zenith angle data measured at the beginning of each frame. The strength of the DAR is largely controlled by the zenith angle, but other atmospheric parameters still play a role. We utilize the \texttt{Starlink} function \texttt{sla\_refco} \citep{Currie_2014} to calculate the coefficients for calculating the ADAR and CDAR. Additionally, using the function \texttt{AltAz} from \texttt{astropy} \citep{astropy_2013} and the Gaia coordinates, we determine the zenith angle of each star in the frame based on the observation time and the geographic coordinates of the Shane and Keck telescopes. We calculate an ``effective wavelength'' ($\lambda_\text{eff}$) using Equation 8 from \citep{Gubler_1998}, integrating over H band. For the reference stars, we assume a temperature of \SI{7000}{\kelvin} as they are likely to be field FG dwarfs (the resulting DAR corrections are relatively insensitive to this assumption). To calculate the effective wavelength for the brown dwarf we directly integrate spectra from 2MASS 0559-14, a T5 dwarf, using data from \cite{McLean_2003}. This calculation yields effective wavelengths of \SI{1.620}{\micro\meter} and \SI{1.635}{\micro\meter} for the brown dwarf and reference star, respectively, in SHARCS H band. Similarly, we calculate \SI{2.110}{\micro\meter} and \SI{2.134}{\micro\meter} for the brown dwarf and reference star, respectively, in NIRC2 Ks band. Note that at the L-T transition, the color reversal makes it possible for the brown dwarf to have a shorter effective wavelength than the reference star in H and Ks bands \citep{Dupuy_2012}. 

For the Shane observations the standard deviations of the set of ADAR values are \SI{2.074}{\mas}, \SI{1.735}{\mas} and the standard deviations of the set of CDAR values are \SI{0.389}{\mas}, \SI{0.144}{\mas} for WISE1901 and WISE2154, respectively. Similarly, for the Keck observations we measure the standard deviation of ADAR values of our targets to be \SI{1.923}{\mas} and \SI{1.527}{\mas} and the standard deviation of the CDAR values of our targets to be \SI{0.251}{\mas} and \SI{0.008}{\mas} for WISE1901 and WISE2154, respectively. The perturbative effects of both ADAR and CDAR on our positions are small relative to our positional error bars and so we do not account for ADAR and CDAR explicitly in our position measurements. However, we add these terms in quadrature to our existing error estimation.

\subsection{Monte Carlo Analysis} \label{subsubsec:mca}

To derive errors bars on our parallax and proper motion measurements, we perform a Monte Carlo simulation of our coordinate transformation, alignment scheme, and parallax and proper motion fitting procedure. This MC analysis uses our best estimate of the total error on the brown dwarf position as described below.

\begin{equation}
\sigma^2_\text{tot} = \sigma^2_\text{pos} + \sigma^2_\text{sys} + \sigma^2_\text{ADAR} + \sigma^2_\text{CDAR}
\label{eq:err}
\end{equation}

Adding the positional error from the brown dwarf data, the systematic overfitting error from M92 data, and the error of ADAR and CDAR in quadrature (Equation \ref{eq:err}), we measure total errors. For the Shane data, we measure errors ($\sigma_x$, $\sigma_y$) of (\SI{11.428}{\mas}, \SI{10.383}{\mas}) for WISE1901 and (\SI{10.599}{\mas}, \SI{7.666}{\mas}) for WISE2154. For the Keck data, we measure errors ($\sigma_x$, $\sigma_y$), of (\SI{11.397}{\mas}, \SI{10.350}{\mas}) for WISE1901 and (\SI{10.566}{\mas}, \SI{7.620}{\mas}) for WISE2154.

For the Monte Carlo analysis we perturb the position of each stellar object (including the brown dwarf) in the frame by sampling Gaussian distributions with width, $\sigma_\text{tot}$, equal to our previously determined total error. For each Monte Carlo iteration, we follow the same alignment and parallax fitting procedure as is performed for the unperturbed data. The positional perturbation is injected at the beginning of the within-epoch alignment step. Sampling our distribution 10000 times yields Gaussian posterior distributions for parallax and proper motion for each of our brown dwarf targets. We then measure the error on each of our model parameters by calculating the standard deviation of the posterior functions.

\section{Results} \label{sec:res}

\begin{deluxetable*}{ccccccc}
\tabletypesize{\normalsize}
\tablecaption{List of Shane and Keck Observations}
\tablehead{
 \colhead{Date} & \colhead{Start Time} & \colhead{Instrument} & \colhead{Band} & \colhead{No. Frames} & \colhead{Integration Time} & \colhead{Mean FWHM} \\
 \colhead{UTC} & \colhead{UTC} & \colhead{} & \colhead{} & \colhead{} & \colhead{seconds} & \colhead{px}
}
\startdata
\multicolumn{7}{c}{\textbf{WISE1901}} \\
\hline
2015-06-10 & 09:51:04.636 & SHARCS, Shane & H & 2 & 120 & 13.423\\
2015-08-08 & 08:49:16.143 & SHARCS, Shane & H & 6 & 120 & 14.677\\
2015-08-30 & 06:40:27.708 & SHARCS, Shane & H & 10 & 120 & 15.199\\
2015-10-04 & 04:14:26.476 & SHARCS, Shane & H & 4 & 120 & 14.953\\
2018-08-26 & 05:42:32.297 & SHARCS, Shane & H & 6 & 120 & 13.164\\
2019-07-21 & 06:48:22.497 & SHARCS, Shane & H & 7 & 120 & 15.952\\
2019-08-17 & 07:24:33.809 & SHARCS, Shane & H & 9 & 120 & 13.284\\
2014-08-06 & 10:11:55.01 & NIRC2, Keck & Ks & 4 & 120 & 2.962\\
2014-09-08 & 05:59:49.44 & NIRC2, Keck & Ks & 8 & 120 & 3.124\\
2015-06-04 & 08:57:52.14 & NIRC2, Keck & Ks & 4 & 120 & 3.056\\
2015-06-30 & 08:10:40.09 & NIRC2, Keck & Ks & 4 & 120 & 3.723\\
\hline
\multicolumn{7}{c}{\textbf{WISE2154}} \\
\hline
2014-11-08 & 09:29:18.811 & SHARCS, Shane & H & 12 & 120 & 11.863\\
2015-08-08 & 09:29:18.811 & SHARCS, Shane & H & 10 & 120 & 15.190\\
2015-08-30 & 09:06:54.778 & SHARCS, Shane & H & 12 & 120 & 15.131\\
2015-10-04 & 05:17:06.431 & SHARCS, Shane & H & 6 & 120 & 15.070\\
2016-06-20 & 10:59:51.252 & SHARCS, Shane & H & 18 & 120 & 13.203\\
2016-09-15 & 05:49:23.903 & SHARCS, Shane & H & 8 & 120 & 14.747\\
2017-08-05 & 10:49:42.324 & SHARCS, Shane & H & 13 & 120 & 14.539\\
2018-08-26 & 09:05:23.968 & SHARCS, Shane & H & 3 & 120 & 16.170\\
2014-08-06 & 10:26:45.81 & NIRC2, Keck & Ks & 3 & 120 & 5.553\\
2014-09-08 & 08:25:23.85 & NIRC2, Keck & Ks & 11 & 120 & 4.743\\
\enddata
\tablecomments{All of our Keck data was taken in the span of about ten months. Shane data was taken over the course of five years until ShaneAO temporarily went offline in 2019 due to low laser guide star return. Furthermore, Shane data was typically obtained around the same time of the year due to poorer weather in the winter and spring months.}
\label{tab:obs}
\end{deluxetable*}

\subsection{WISE1901} \label{subsec:wise1901}

\begin{figure*}[ht]
\centering
\includegraphics[height=0.6\columnwidth]{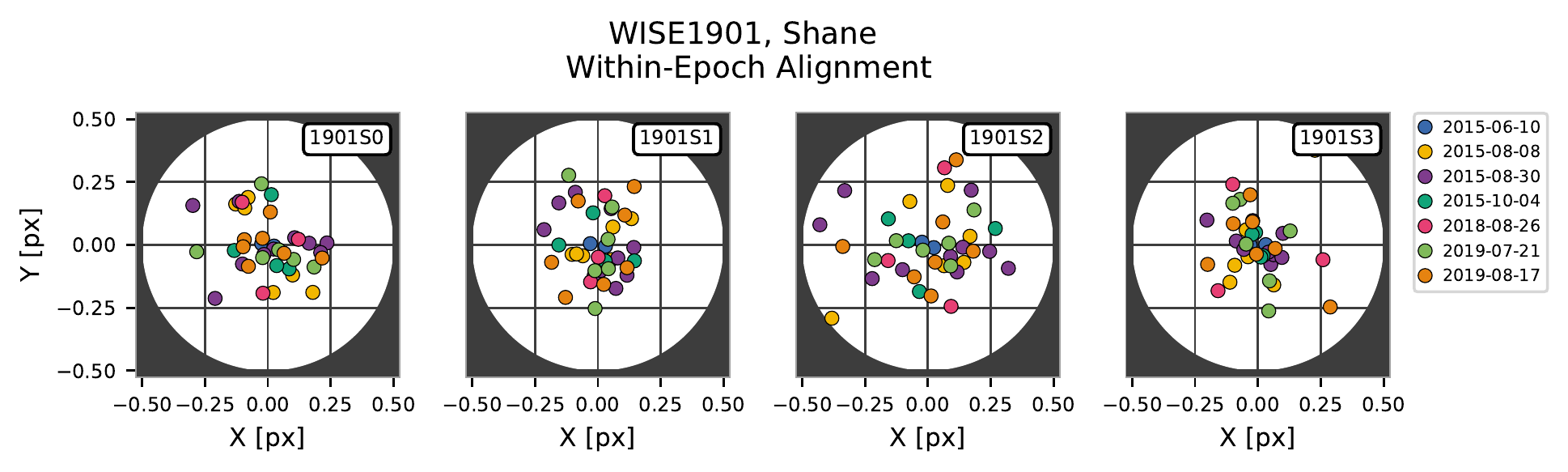}
\caption{Eight epochs of residuals from the alignment of frames within-epoch for WISE1901 Shane data. We see relatively good dispersion and the points are mostly aligned. The third object was our faintest source, as a result, it has larger residuals than the rest of the objects. The standard deviations for these residuals, in units of Shane pixels, are: ($\sigma_x$,$\sigma_y$) = (\SI{0.128}{\pixel}, \SI{0.115}{\pixel}), (\SI{0.095}{\pixel}, \SI{0.130}{\pixel}), (\SI{0.179}{\pixel}, \SI{0.141}{\pixel}), (\SI{0.108}{\pixel}, \SI{0.127}{\pixel}) for the four objects in the frame, in order.}
\label{fig:ea_shane_WISE1901}
\end{figure*}

\begin{figure*}[ht]
\centering
\includegraphics[height=0.6\columnwidth]{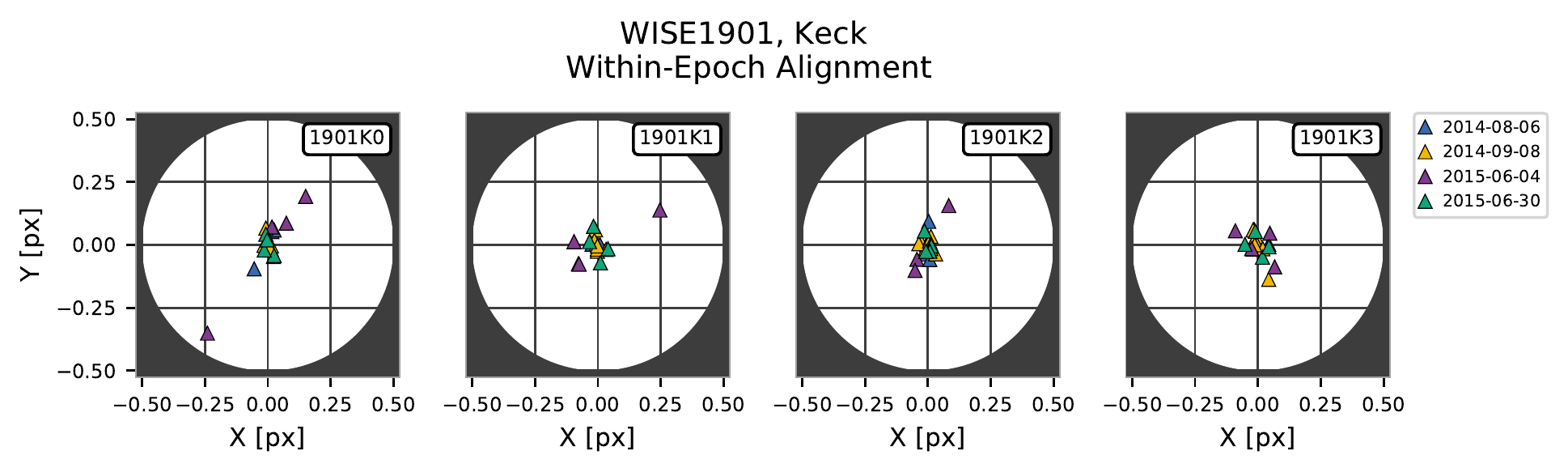}
\caption{Four epochs of residuals from the alignment of frames within-epoch for WISE1901 Keck data. Keck data was much better resolved than Shane data, even for the faint object. The standard deviations for these residuals, in units of Keck pixels, are: ($\sigma_x$,$\sigma_y$) = (\SI{0.068}{\pixel}, \SI{0.100}{\pixel}), (\SI{0.066}{\pixel}, \SI{0.049}{\pixel}), (\SI{0.028}{\pixel}, \SI{0.054}{\pixel}), (\SI{0.036}{\pixel}, \SI{0.048}{\pixel}) for the four objects in the frame, in order.}
\label{fig:ea_keck_WISE1901}
\end{figure*}

\begin{figure*}[ht]
\centering
\includegraphics[height=0.6\columnwidth]{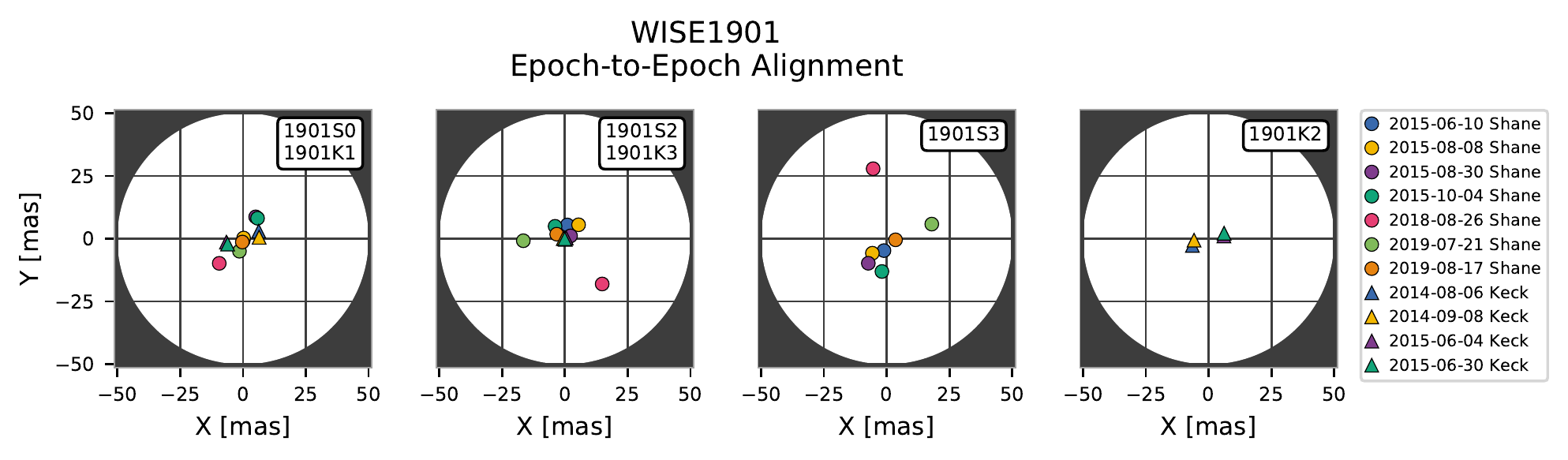}
\caption{Residuals from Shane and Keck data of epoch-to-epoch alignment for WISE1901 in units of Shane pixels. Shane and Keck are not on the ICRS yet and their fits are independent one another. The eight Shane epoch residuals are tightly clustered when aligning between epochs. The Keck residuals shows clustering for the pairs of data taken in 2014 and 2015, but are overall still tightly clustered. Across all objects, we measure overall standard deviations of ($\sigma_x$,$\sigma_y$) = (\SI{7.433}{\mas}, \SI{9.291}{\mas}) for the Shane data and ($\sigma_x$,$\sigma_y$) = (\SI{5.104}{\mas}, \SI{1.573}{\mas}) for the Keck data.}
\label{fig:e2e_WISE1901}
\end{figure*}

\begin{figure*}[ht]
\centering
\includegraphics[height=0.6\columnwidth]{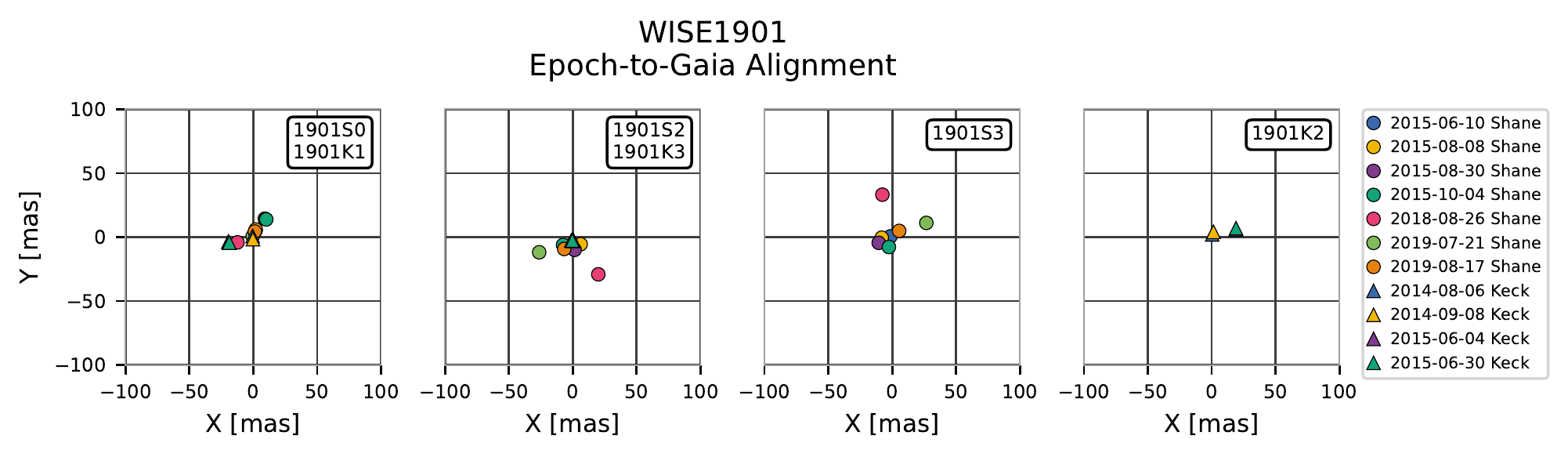}
\caption{Residuals from WISE1901 epoch-to-Gaia alignment for both Shane and Keck. The final alignment of the Shane data to Gaia shows somewhat large residuals, but the large number of epochs helps smooth out the final fit. Across all objects, we measure overall standard deviations of ($\sigma_x$,$\sigma_y$) = (\SI{11.056}{\mas}, \SI{12.144}{\mas}) for the Shane data and ($\sigma_x$,$\sigma_y$) = (\SI{11.023}{\mas}, \SI{3.731}{\mas}) for the Keck data.}
\label{fig:e2g_WISE1901}
\end{figure*}

\begin{figure*}[ht]
\centering
\includegraphics[width=1.4\columnwidth]{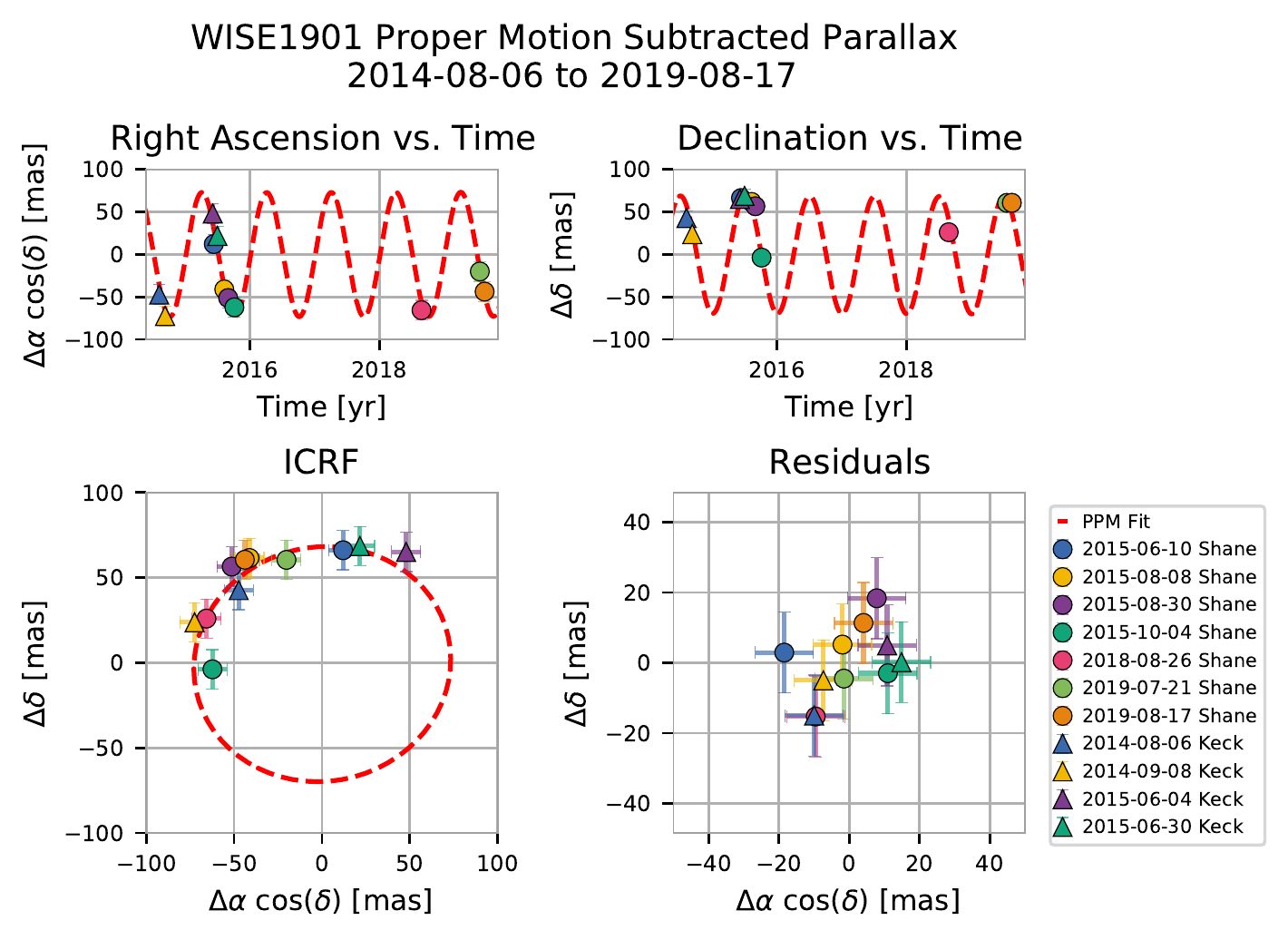}
\caption{Final parallax and proper motion fit for WISE1901. Note that the data was generally taken around the same time of year every year. The residuals have a standard deviation of ($\sigma_x$,$\sigma_y$) = (\SI{10.129}{\mas}, \SI{9.711}{\mas}). The reduced chi-squared of the fit is $\chi^2_\nu = 1.152$.}
\label{fig:ppm_WISE1901}
\end{figure*}

The WISE1901 field as seen by Shane and Keck do not share the same set of reference stars. In particular, there is a very bright star (1901S3) in the Shane frame that is just outside the field in the Keck frame. For the Shane data we include this bright star as an astrometric reference while for the Keck data we include a fainter nearby object (1901K2) instead. The very bright star is near the edge of the frame and a nearby fainter star also influences the PSF, resulting in poorer positional precision. The Shane position measurement error was therefore significantly higher for this target. For WISE1901, we measure \wiseBParallax{}, \wiseBProperMotionRA{}, and \wiseBProperMotionDec{}. Figures \ref{fig:ea_shane_WISE1901} and \ref{fig:ea_keck_WISE1901} show the within-epoch residuals, Figure \ref{fig:e2e_WISE1901} show the epoch-to-epoch residuals, Figure \ref{fig:e2g_WISE1901} shows the epoch-to-Gaia residuals, and Figure \ref{fig:ppm_WISE1901} shows the parallax and proper motion fit. The overall residuals for the Figure \ref{fig:ppm_WISE1901} are clustered with no clear bias, but shows the low precision due to the bright reference star in the Shane data. Compared to \cite{Kirkpatrick_2020}, our parallax is within $0.7\sigma$ of their measured value of $\pi = \SI{67.3\pm3.4}{\mas}$. Their proper motion values of $\mu_\alpha = \SI{122.9\pm1.4}{\mas\per\year}$ and $\mu_\delta = \SI{405.7\pm1.4}{\mas\per\year}$ are within $1.7\sigma$ and $1.0\sigma$ of our values, respectively. The proper motion in right ascension for WISE1901 is the largest deviation in our data set, but is still within a $2\sigma$. The chance of this result occurring is small, but not too large to be totally improbable. Therefore, we still find our results to be consistent with \cite{Kirkpatrick_2020}.

\subsection{WISE2154} \label{subsec:wise2154}

\begin{figure*}[ht]
\centering
\includegraphics[height=0.6\columnwidth]{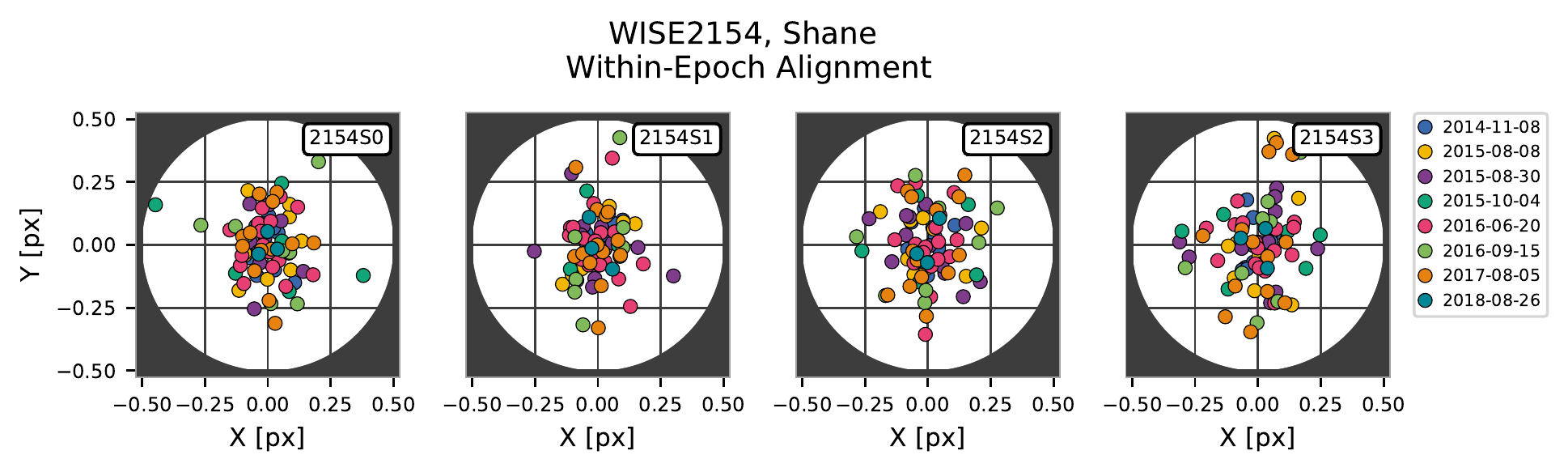}
\caption{Eight epochs of residuals from the alignment of frames within-epoch for WISE2154 Shane data. This data is our cleanest set of data. The standard deviations for these residuals, in units of Shane pixels, are: ($\sigma_x$,$\sigma_y$) = (\SI{0.104}{\pixel}, \SI{0.123}{\pixel}), (\SI{0.082}{\pixel}, \SI{0.126}{\pixel}), (\SI{0.105}{\pixel}, \SI{0.134}{\pixel}), (\SI{0.112}{\pixel}, \SI{0.154}{\pixel}) the four objects in the frame, in order.}
\label{fig:ea_shane_WISE2154}
\end{figure*}

\begin{figure*}[ht]
\centering
\includegraphics[height=0.6\columnwidth]{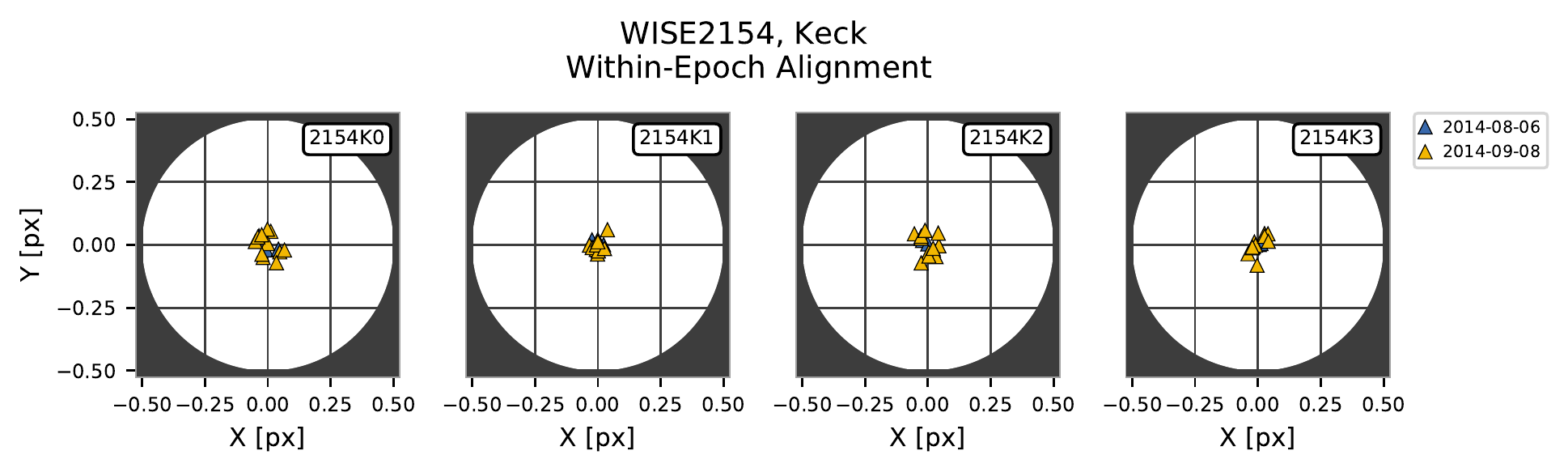}
\caption{Two epochs of residuals from the alignment of frames within-epoch for WISE2154 Keck data. The standard deviations for these residuals, in units of Keck pixels, are: ($\sigma_x$,$\sigma_y$) = (\SI{0.034}{\pixel}, \SI{0.039}{\pixel}), (\SI{0.019}{\pixel}, \SI{0.024}{\pixel}), (\SI{0.028}{\pixel}, \SI{0.038}{\pixel}), (\SI{0.025}{\pixel}, \SI{0.031}{\pixel}) for the four objects in the frame, in order.}
\label{fig:ea_keck_WISE2154}
\end{figure*}

\begin{figure*}[ht]
\centering
\includegraphics[height=0.6\columnwidth]{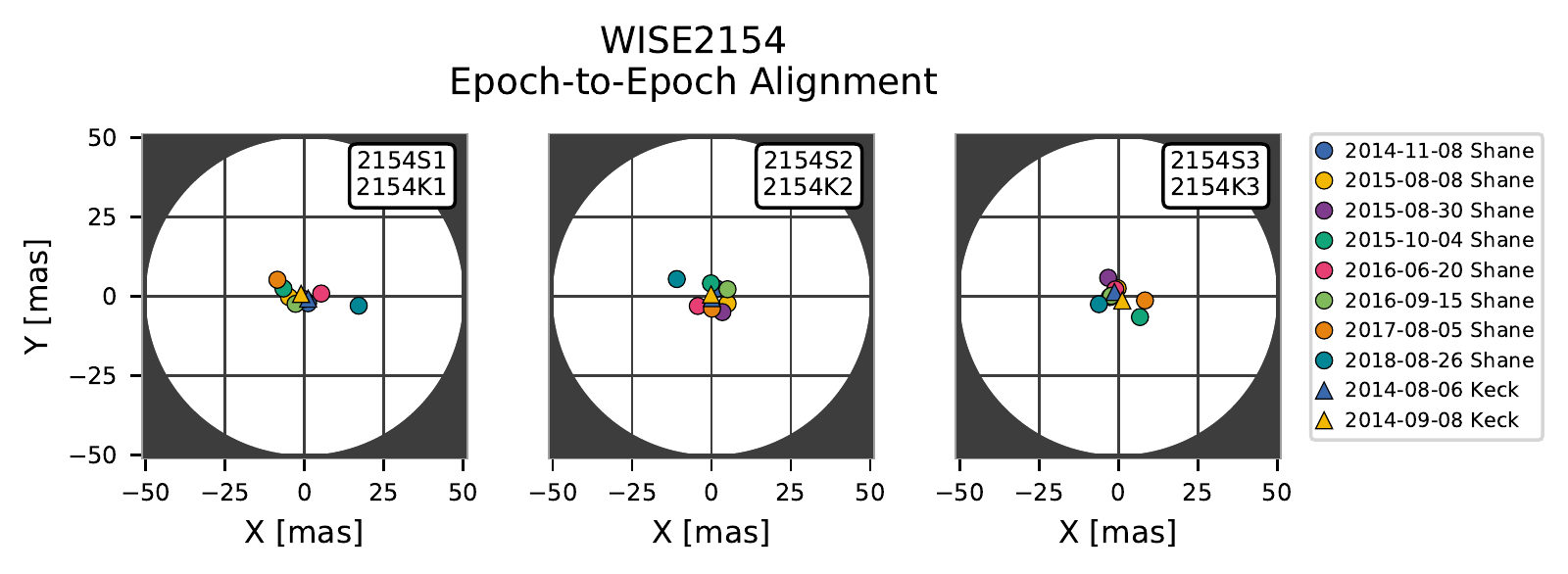}
\caption{Residuals from Shane and Keck data of epoch-to-epoch alignment for WISE2154 in units of Shane pixels. Shane and Keck are not on the ICRS yet and their fits are independent one another. Across all objects, we measure overall standard deviations of ($\sigma_x$,$\sigma_y$) = (\SI{5.952}{\mas}, \SI{3.314}{\mas}) for the Shane data and ($\sigma_x$,$\sigma_y$) = (\SI{0.936}{\mas}, \SI{0.948}{\mas}) for the Keck data.}
\label{fig:e2e_WISE2154}
\end{figure*}

\begin{figure*}[ht]
\centering
\includegraphics[height=0.6\columnwidth]{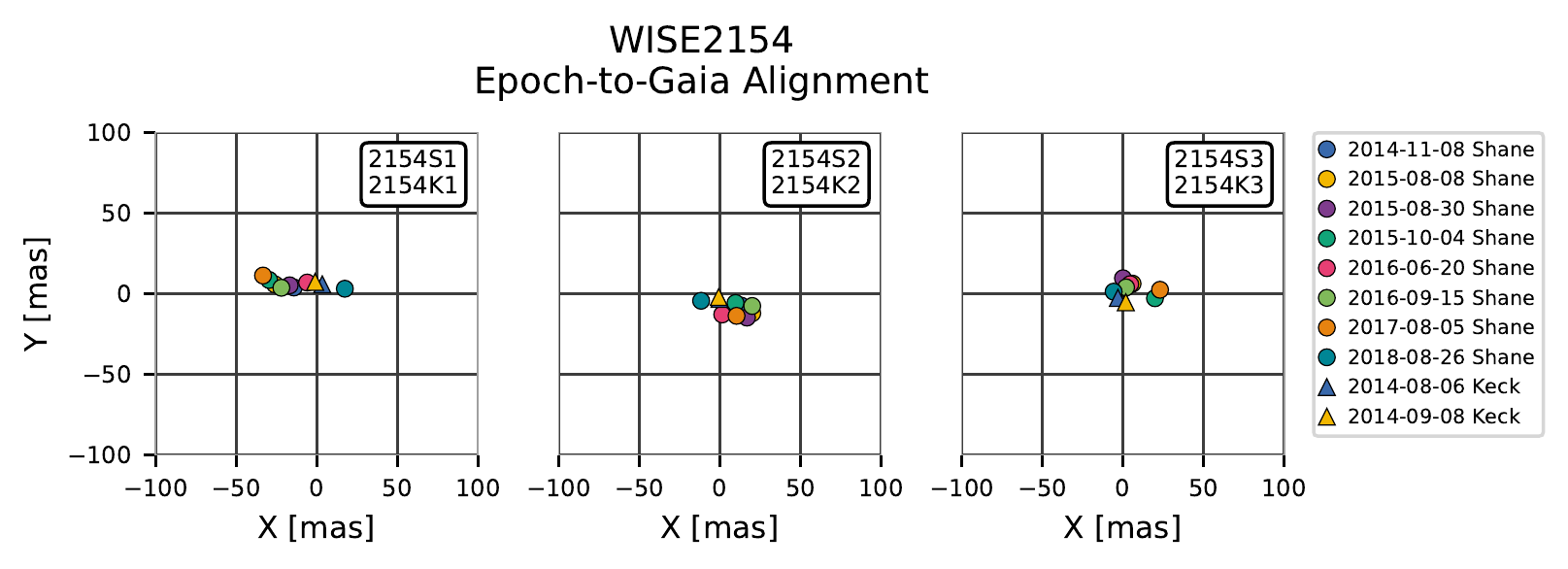}
\caption{Residuals from WISE2154 epoch-to-Gaia alignment for both Shane and Keck. Shane epoch-to-Gaia fits are slightly offset from the reference stars. Across all objects, we measure overall standard deviations of ($\sigma_x$,$\sigma_y$) = (\SI{16.546}{\mas}, \SI{7.805}{\mas}) for the Shane data and ($\sigma_x$,$\sigma_y$) = (\SI{2.055}{\mas}, \SI{4.679}{\mas}) for the Keck data.}
\label{fig:e2g_WISE2154}
\end{figure*}

\begin{figure*}[ht]
\centering
\includegraphics[width=1.4\columnwidth]{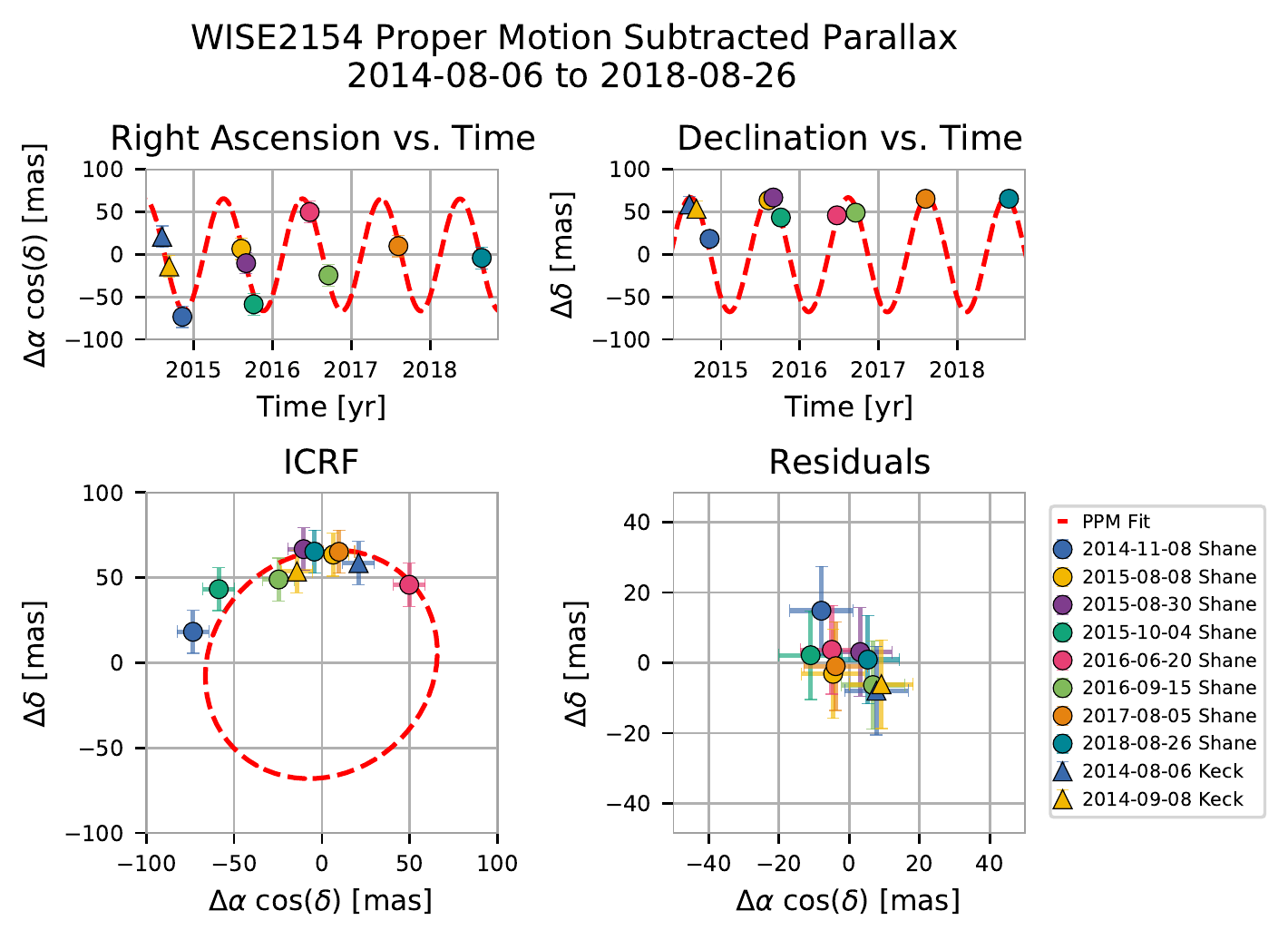}
\caption{Final parallax and proper motion fit for WISE2154. This target has has the smallest residuals of all of our targets. In particular, WISE2154 is well sampled throughout the periodic motion of the brown dwarf. The residuals have a standard deviation of ($\sigma_x$,$\sigma_y$) = (\SI{6.860}{\mas}, \SI{6.316}{\mas}). The reduced chi-squared of the fit is $\chi^2_\nu = 1.338$.}
\label{fig:ppm_WISE2154}
\end{figure*}

\begin{deluxetable*}{ccccccc}
\label{tab:results}
\tabletypesize{\normalsize}
\tablecaption{Final parallax and proper motion Fit Measurements}
\tablehead{
\colhead{Target} & \colhead{RA} & \colhead{Dec} & \colhead{PM RA} & \colhead{PM Dec} & \colhead{Parallax} & \colhead{$\chi^2_\nu$}\\
\colhead{at 2015.5} & \colhead{deg} & \colhead{deg} & \colhead{mas/yr} & \colhead{mas/yr} & \colhead{mas} & \colhead{}
}
\decimals
\startdata
\wiseBTableRow
\wiseCTableRow
\enddata
\end{deluxetable*}

We present a total of seven Shane epochs and two Keck epochs for WISE2154. In both the Shane and Keck data, we utilize the same set of three reference stars. For WISE2154, we measure \wiseCParallax{}, \wiseCProperMotionRA{}, and \wiseCProperMotionDec{}. Figures \ref{fig:ea_shane_WISE2154} and \ref{fig:ea_keck_WISE2154} show the within-epoch residuals, Figure \ref{fig:e2e_WISE2154} show the epoch-to-epoch residuals, Figure \ref{fig:e2g_WISE2154} shows the epoch-to-Gaia residuals, and Figure \ref{fig:ppm_WISE2154} shows the parallax and proper motion fit. The parallax and proper motion fit for this target has small, unbiased residuals and more even temporal sampling compared to the previous target. \cite{Kirkpatrick_2020} measures $\pi = \SI{71.0\pm2.3}{\mas}$, $\mu_\alpha = \SI{-164.9\pm0.7}{\mas\per\year}$, $\mu_\delta = \SI{-465.0\pm0.7}{\mas\per\year}$, corresponding to within $0.1\sigma$, $0.6\sigma$, and $0.5\sigma$ of our results, respectively. The \cite{Kirkpatrick_2020} measurements are all consistent with ours, to within $1\sigma$.

\section{Conclusion} \label{sec:conc}
In this paper we present parallax and proper motion measurements of two T dwarfs using combined observations from the Shane and Keck telescopes spanning five years. Results are summarized in Table \ref{tab:results}. All observations utilized laser guide star adaptive optics systems on these telescopes. We utilize a Monte Carlo simulation of our sparse-reference star procedure to quantify our parallax and proper motion errors, using our best estimates of both the random and systematic portions of the astrometric errors. We constrain the parallax and proper motion components of WISE1901 to be \wiseBParallax{}, \wiseBProperMotionRA{}, \wiseBProperMotionDec{}. For WISE2154, we measure \wiseCParallax{}, \wiseCProperMotionRA{}, \wiseCProperMotionDec{}. All our results are consistent within $2\sigma$, with 4 of 6 values within $1\sigma$, of previously published five-parameter astrometric solutions in the literature. 

This study leverages five-parameter parallax and proper motion solutions for our reference stars from Gaia DR2. Use of Gaia data allow us to place our measurements on the ICRS astrometric reference frame as well as mitigate errors due to unknown motions of background reference stars. For these reasons, we are able to constrain the parallax and proper motions of these two brown dwarfs using only three Gaia reference stars per target.

These two targets are the first of a larger sample of T dwarfs being astrometrically monitored. With the data analysis pipeline established, future publications will include the remaining targets once a minimum of six epochs per target are reached. Distances to the objects and luminosities may be derived for these data, and these measurements may be used as part of larger samples to calibrate the lowest-mass end of stellar formation models.

\section{Acknowledgements} \label{sec:ack}

We acknowledge all members of the ShaneAO commissioning team for assistance in obtaining M92 images as well as Sasha Safonova and Don Pham for assistance with data reduction. This work performed under the auspices of the U.S. Department of Energy by Lawrence Livermore National Laboratory under Contract DE-AC52-07NA27344 with release number LLNL-JRNL-821003. Some of the data presented herein were obtained at the W. M. Keck Observatory, which is operated as a scientific partnership among the California Institute of Technology, the University of California and the National Aeronautics and Space Administration. The Observatory was made possible by the generous financial support of the W. M. Keck Foundation. The authors wish to recognize and acknowledge the very significant cultural role and reverence that the summit of Maunakea has always had within the indigenous Hawaiian community. We are most fortunate to have the opportunity to conduct observations from this mountain. This work has made use of data from the European Space Agency (ESA) mission {\it Gaia} (\url{https://www.cosmos.esa.int/gaia}), processed by the {\it Gaia} Data Processing and Analysis Consortium (DPAC, \url{https://www.cosmos.esa.int/web/gaia/dpac/consortium}). Funding for the DPAC has been provided by national institutions, in particular the institutions participating in the {\it Gaia} Multilateral Agreement. This research made use of Astropy,\footnote{http://www.astropy.org} a community-developed core Python package for Astronomy \citep{astropy_2013, astropy_2018}. The Starlink software \citep{Currie_2014} is currently supported by the East Asian Observatory.

\software{astropy \citep{astropy_2013, astropy_2018}, astroscrappy \citep{McCully_2018}, Python \citep{Rossum_1995}, StarFinder \citep{Diolati_2000}, Starlink \citep{Currie_2014}}

\clearpage
\bibliography{ref}{}
\bibliographystyle{aasjournal}

\end{document}